\documentclass[referee]{aa}  

\usepackage{graphicx}
\usepackage{txfonts}
\usepackage{color}

\usepackage{multirow}  

\begin{document} 

\title{CIELO-RGS: a catalogue of soft X-ray ionized emission lines}

\author{Junjie Mao\inst{\ref{inst1},\ref{inst_strath}}
  \and J.~S. Kaastra\inst{\ref{inst1},\ref{inst5}}
  \and M.~Guainazzi\inst{\ref{inst2}}
  \and R.~Gonz\'alez-Riestra\inst{\ref{inst3}}
  \and M.~Santos-LLe\'o\inst{\ref{inst3}}
  \and P.~Kretschmar\inst{\ref{inst3}}
  \and V.~Grinberg\inst{\ref{inst4}}
  \and E.~Kalfountzou\inst{\ref{inst3}}
  \and A.~Ibarra\inst{\ref{inst3}}
  \and G.~Matzeu\inst{\ref{inst3}}
  \and M.~Parker\inst{\ref{inst3}}
  \and P.~Rodr\'iguez-Pascual\inst{\ref{inst3}}
          }

\institute{SRON Netherlands Institute for Space Research, Sorbonnelaan 2, 3584 CA, Utrecht, The Netherlands\label{inst1}
  \and Department of Physics, University of Strathclyde, Glasgow G4 0NG, UK\label{inst_strath}
  \and Leiden Observatory, Leiden University, PO Box 9513,
   2300 RA Leiden, the Netherlands\label{inst5}
  \and ESA European Space Research and Technology Centre (ESTEC), Keplerlaan 1, 2201 AZ, Noordwijk, The Netherlands\label{inst2}
  \and European Space Agency (ESA), European Space Astronomy Centre (ESAC), Villanueva de la Cañada, E-28691 Madrid, Spain\label{inst3}
  \and Institut f\"ur Astronomie und Astrophysik, Universit\"at T\"ubingen, Sand 1, 72076 T\"ubingen, Germany\label{inst4}
             }

   \date{}

\abstract
{High-resolution X-ray spectroscopy has advanced our understanding of the hot Universe by revealing physical properties like kinematics, temperature, and abundances of the astrophysical plasmas. Despite the technical and scientific achievements, the lack of scientific products at a level higher than count spectra is hampering full scientific exploitation of high-quality data.  This paper introduces the Catalogue of Ionized Emission Lines Observed by the Reflection Grating Spectrometer (CIELO-RGS) onboard the \textit{XMM-Newton} space observatory.}
{The CIELO-RGS catalogue aims to facilitate the exploitation of emission features in the public RGS spectra archive, in particular, to perform the correlation between X-ray spectral diagnostics parameters with measurements at other wavelengths. This paper focuses on the methodology of catalogue generation, describing the automated line detection algorithm.}
{A moderate sample ($\sim2400$ observations) of high-quality RGS spectra available at \textit{XMM-Newton} Science Archive is used as our starting point. A list of potential emission lines is selected based on a multi-scale peak detection algorithm in a uniform and automated way without prior assumption on the underlying astrophysical model. The candidate line list is validated via spectral fitting with simple continuum and line profile models. We also compare the catalogue content with published literature results on a small number of exemplary sources.}
{We generate a catalogue of emission lines ($1.2\times10^4$) detected in $\sim1600$ observations toward stars, X-ray binaries, supernovae remnants, active galactic nuclei, and groups and clusters of galaxies. For each line, we report the observed wavelength, broadening, energy and photon flux, and equivalent width, etc.}
{} 
 
             
\titlerunning{CIELO-RGS}
\authorrunning{J.Mao et al.}
  
\maketitle

\section{Introduction}
\label{sect_introduction}

High-resolution X-ray spectroscopy has the potential to reveal the physics of the plasma state that accounts for much of the material in the observable universe. The X-ray band contains the inner-shell transitions of all abundant elements from C through N, O, Ne, Mg, Si, S, Ca, Fe and Ni. X-ray spectra of cosmic sources typically consist of a mixture of ionic or atomic emission or absorption lines and electron continuum components. Observed spectra can be modeled in order to identify the underlying physics and dynamics of the plasma concerned. They often depart from the ideal conditions of equilibrium and spatial uniformity. It is not coincidental that the availability of high-resolution data in the X-ray domain at the beginning of the current millennium has triggered important development in the modeling of X-ray emitting plasma \citep{behar01,hitomi18}, in the collisionally ionized as well as in the photo-ionized regime.

A giant leap in this field was achieved with the launch of \textit{Chandra} and \textit{XMM-Newton} carrying high-sensitivity, high-resolution grating systems. In particular, the Reflection Grating Spectrometer (RGS; \cite{denherder01}) on board \textit{XMM-Newton} \citep{jansen01} consists of an array of reflection gratings that diffract the light focused by two of the X-ray telescopes to an array of dedicated Charge-Coupled Device detectors. The RGS achieves high resolving power (150 to 800) over an energy range from ~0.35 to 2 keV ($\simeq$6 to 35\AA, in the first spectral order). Thanks to the unprecedented collecting area of the \textit{XMM-Newton} telescopes, the RGS is the most sensitive high-resolution X-ray detector currently in operation. Its performance and excellent calibration quality \citep{devries15} have permitted to obtain breakthrough advancements in astrophysical fields as diverse as cool and active stars, X-ray binaries, galaxy clusters, and Active Galactic Nuclei (AGN), to mention just a few \citep{Kahn02}.

Despite these outstanding technical and scientific achievements, the full scientific exploitation of the plasma diagnostic information enclosed in RGS spectroscopic data is still partly hampered by the lack of scientific products at a level higher than count spectra. Astronomers interested in extracting plasma diagnostics from RGS spectra must analyze them through the “forward-folding technique” requiring specialized software, potentially cumbersome when applied to spectra with good signal-to-noise ratio.

This paper describes \textit{CIELO-RGS}\footnote{\textit{Catalogue of Ionized Emission Lines Observed by the RGS}} ("CIELO" thereafter), a catalogue of emission lines detected by the RGS instrument over more than 15 years of \textit{XMM-Newton} science operations. To our knowledge CIELO is the first published catalogue of this kind in high-energy astrophysics. The paper is organized as follows. Sect.~\ref{sect_reduction} describe the procedure followed to reduce and select the optimal sample of RGS the spectra over which the CIELO catalogueue has been produced. Sect.~\ref{sect_catalogue} describes the algorithm employed to select a sample of candidate emission lines on each RGS spectrum, later validated through spectral fitting. The catalogue structure is described in Sect.~\ref{sect_structure}. A selection of a few preliminary science results of the catalogue as a whole are presented
in Sect.~\ref{sect_results}, followed by our conclusions (Sect.~\ref{sect_discussion}. Future papers will discuss the application of CIELO to a wide range of astrophysical contexts and problems.

The CIELO catalog, data and source codes used for creating the figures in this paper can be found at Zenodo\footnote{http://doi.org/10.5281/zenodo.2577859}. 

\section{Spectra selection and reduction}
\label{sect_reduction}

The CIELO Catalogue is based on a subset of the RGS spectra available in the \textit{XMM-Newton} Science Archive (XSA).

This subset has been taken from the work of \citet{ben14}. These authors examined all the RGS data available as of 1st December 2014 in order to identify those that are scientifically useful.  

The method devised was based on two properties. The first one was an estimation of the signal-to-noise ratio of the spectrum, for which a minimum value of 10 was required.  The second one was the spatial profile of the spectrum along the cross-dispersion direction of the spectral image. This profile was fitted with a Gaussian function.  For the spectrum to be considered useful, the fitting parameters needed to be well determined and within some pre-defined range (e.g. the width of the fitted Gaussian could not be narrower than the instrumental spatial resolution, the maximum had to be close to the center of the field-of-view).

Out of the 8500 observations studied, 5000 (59\%) passed these two acceptance criteria. These 5000 observations were further sub-divided into two categories: ``Top" observations for which both RGS1 and RGS2 spectra were selected (28\% of the initial sample), and "Good" observations, with only RGS1 or RGS2 considered useful (31\% of the initial sample). The sample used here is composed of the 2421 observations classified as ``Top quality" in \citet{ben14}.


The data used in this work have been taken from the \textit{XMM-Newton} Science Archive (XSA). For each of the selected observations, the following data have been downloaded: extracted first order total
(i.e. source+background) spectra, extracted first order background spectra and first order response matrices (that for RGS also include the effective area), for RGS1 and RGS2.

These data have been obtained from the raw Observation Data File (ODF) after automatic processing with the Pipeline Processing System (PPS). The PPS extracts RGS source spectra using the coordinates given in the original observational proposal, and an extraction region with a spatial size of 95\% of the cross-dispersion PSF. The background is extracted from the region between 98\% and 100\% of the cross-dispersion PSF. A region including 95\% of the pulse-height distribution is used in the extraction of both source and background spectra.

\section{Catalogue generation}
\label{sect_catalogue}
Sophisticated plasma models are widely used for detailed analysis of emission features in the high-resolution X-ray spectra \citep[e.g.,][]{bro11,gu16,mao18,mao19}. While this approach yields self-consistent physical interpretations (e.g., temperature, density, and abundance) of the spectral features, it requires extensive prior knowledge of the source of interest, including but not limited to the cosmological redshift, line-of-sight galactic absorption, which and how many plasma models to use (collisional ionized, photoionized, non-equilibrium ionization, and charge exchange), and sometimes a multi-component model of the continuum. This approach is not applicable for the analysis of a large sample like CIELO-RGS. 

An alternative approach is to use a phenomenological fit \citep[e.g.,][]{gua07,pin16,psa18} for individual emission features. Typically, the (local) continuum is simply modeled as a (local) power law or spline function. Subsequently, it is common for the investigators to identify the emission features by eye or to scan the spectrum using a Gaussian/delta profile. Each emission features is then modeled with a delta function or a Gaussian function with fixed width. 

The phenomenological analysis is straightforward. It requires no prior knowledge of the source if all the emission lines are narrow and unresolved. When the emission lines are broader than the spectral resolution of the instrument, assuming that all the emission features have identical broadening, then the only prior knowledge needed is the line broadening. 

\begin{figure}
\centering
\includegraphics[width=\hsize]{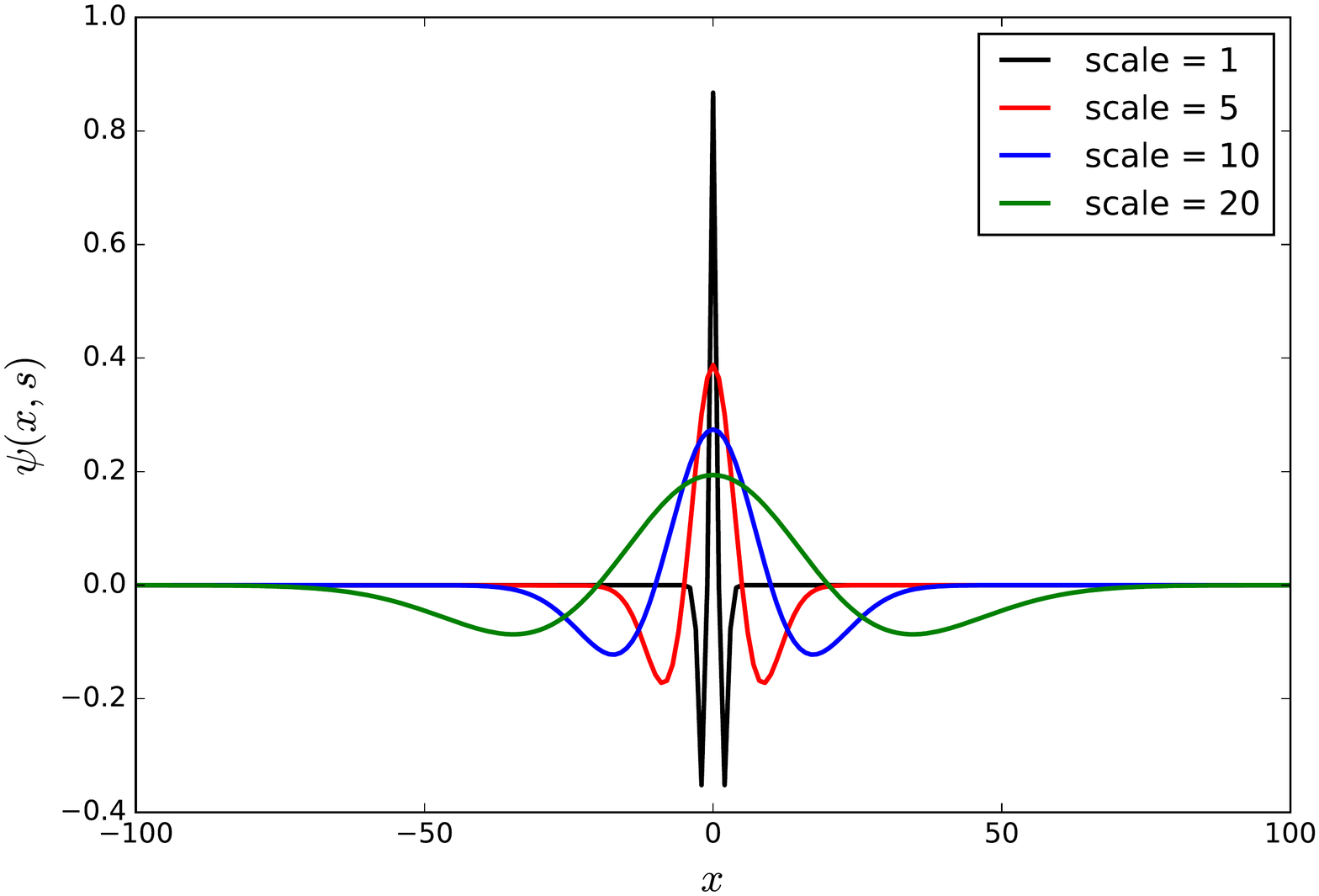}
\caption{The Mexican hat wavelet at different scales. Here we use 201 points (i.e. 200 bins) to describe the wavelet. The corresponding Y-value is placed at the center of the each bin to mimic the real spectra.}
\label{fig:mexh_plot}
\end{figure}

The above issue with conventional phenomenological analysis is tractable with a line detection algorithm based on wavelet transform. Note that wavelet transform analyses have been proven to be suitable for both spectral and imaging analysis in astronomy \citep[e.g.,][]{bur96,sta97,mac13}. 
In particular, the concept of the multi-scale peak detection (MSPD) initially proposed in the chemistry community \citet[][Z15 hereafter]{zha15} was followed for the present work. 

The advantage of MSPD is that it extracts information from high-throughput spectra in a rapid way without any prior knowledge. Strong and weak, broad and narrow, symmetric and asymmetric spectral features on top of a continuum that can be easily detected (Section~\ref{sect_mspd}). Once the number of significant spectral features is determined, the aforementioned phenomenological fit can be performed in a uniform and automated way (Section~\ref{sect_fitting}).  


We should point out that the original MSPD algorithm shown in Z15 needs to be tailored for RGS data analysis. Z15 took advantage of ``ridges" (local maxima, elucidated later in Section~\ref{sect_mspd}), ``valleys" (local minima) and ``zero-crossings" (sign changes) for the peak detection. As MSPD does not estimate/fit the continuum, the presence of continuum breaks in the RGS spectra (e.g., CCD gaps) would lead to artificial ``valleys" and ``zero-crossings". Furthermore, in Z15, the signal and noise levels are defined as the maximum and the 90th percentile within a selected window for each potential peak\footnote{Details can be found in the source code.}. This would not work properly for closely located emission features like He-like triplets in many astrophysical spectra. 

\begin{figure*}
\centering
\includegraphics[width=\hsize, trim={2.0cm 1.0cm 2.0cm 1.0cm}, clip]{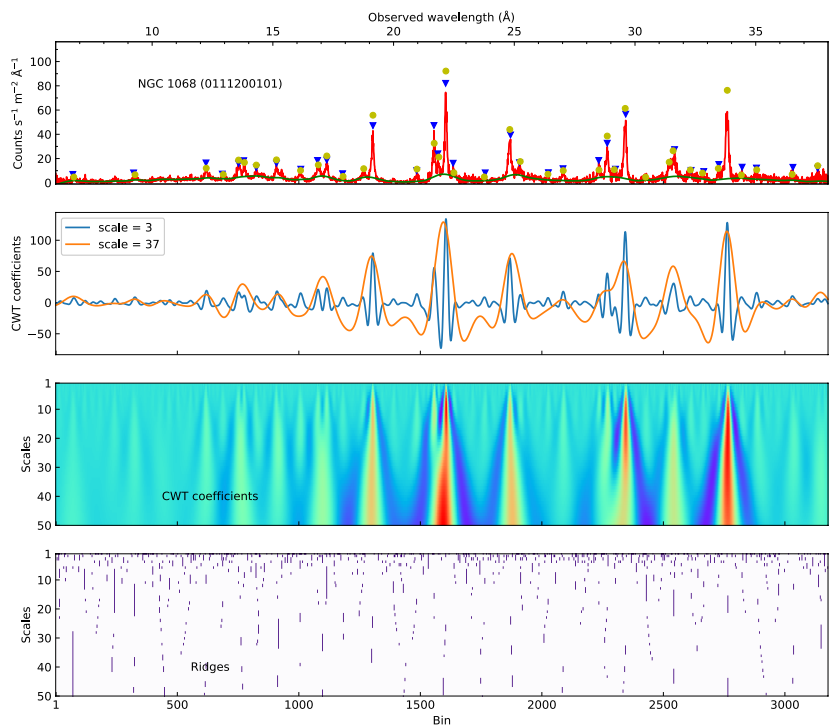}
\caption{The top panel is the observed RGS flux spectrum of NGC\,1068 (ObsID: 0111200101). The RGS1 and RGS2 spectra (in red) are combined. Error bars are not shown for clarity. The blue triangles mark peaks detected by the MSPD algorithm \citep{zha15} with (their definition of) signal to noise ratio $\ge7$. The yellow circles mark peaks detected by the present work at a significance level of $\gtrsim2.7\sigma$. The estimated continuum in green is based on the asymmetrically reweighted penalized least squares algorithm \citep[arPLS,][]{bae15}. The estimated continuum is shown merely for guidance as it is independent of the MSPD algorithm. The second panel gives the coefficients (Y-axis) of the continuous wavelet transform (CWT) of the spectrum at scale$=3$ (blue) and $37$ (orange). The third panel is a color map of the CWT coefficients at different scales (Y-axis) and at each spectral bins (X-axis). The redder the color, the larger the CWT coefficient. The bottom panel presents the so-called ``ridges" across different scales.}
\label{fig:ngc1068_mspd}
\end{figure*}

\subsection{Multi-scale Peak Detection}
\label{sect_mspd}
The present MSPD algorithm flows as follows. It is based on the continuous wavelet transform (CWT) using the Mexican hat wavelet (or Ricker wavelet)
\begin{equation}
    \psi(x, s) = \frac{2}{\sqrt{3~s} \pi^{1/4}}\left(1 - \frac{x^2}{s^2}\right) \exp\left({-\frac{x^2}{2~s^2}}\right)~,
\end{equation}
where $x$ is the number of points used to describe the wavelet and $s$ is the scale of the wavelet. Figure~\ref{fig:mexh_plot} shows an example of the Mexican hat wavelet at some scales.

The input spectrum $S(N)$ is convolved with a set of wavelet functions $\psi(N',~s)$, where $N'$ is the minimum value between the number of bins ($N$) in the input spectrum and ten times the scale ($10\times s$) and the scale ($s$) ranges from 1 to 50. The top panel of Figure~\ref{fig:ngc1068_mspd} shows the observed RGS flux spectrum of NGC\,1068 (ObsID: 0111200101) in the $6-8~\AA$ wavelength range ($\sim3200$ spectral bins). The second panel of Figure~\ref{fig:ngc1068_mspd} presents the results of the continuous wavelet transform (CWT) with $s=3$ in blue and $37$ in orange, respectively. Peaks appear in the observed spectrum yield larger CWT coefficients. The third panel of Figure~\ref{fig:ngc1068_mspd} presents the color-coded CWT coefficients. This can be viewed as the dilation and translation of peaks. The redder (larger) the coefficient, the more prominent the peak. Ridges are introduced as curves of local maxima in the two-dimension (2D) wavelet space, where the rows and columns are scales and bins, respectively. The bottom panel of Figure~\ref{fig:ngc1068_mspd} illustrates ridges across different scales for NGC\,1068. Ridges are denser at smaller scales as the spectrum is nosier.

\begin{figure*}
\centering
\includegraphics[width=\hsize, trim={2.0cm 2.0cm 2.0cm 2.0cm}, clip]{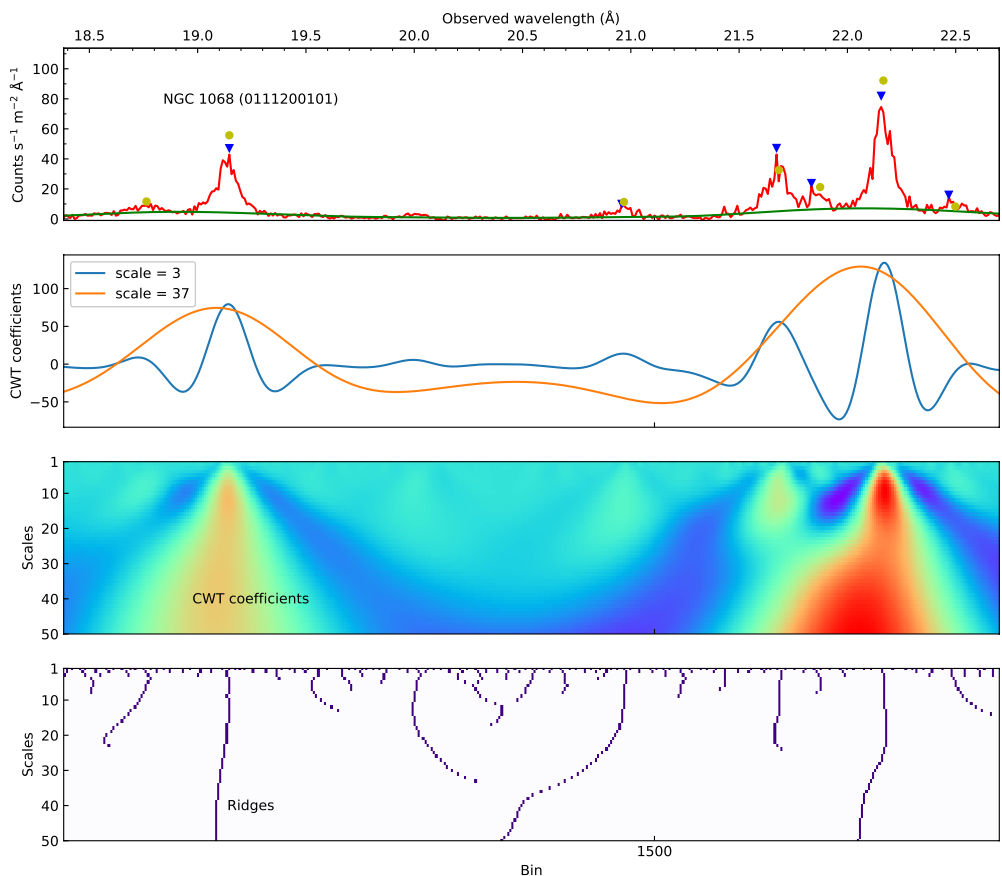}
\caption{Similar to Figure~\ref{fig:ngc1068_mspd} but in the neighbourhood of the \ion{O}{VIII} Ly$\alpha$ line and the \ion{O}{VII} He-like triplets.}
\label{fig:ngc1068_mspd_zoom}
\end{figure*}

A ridge in the wavelet space does not necessarily corresponds to a prominent emission feature in the observed spectrum (e.g. Figure~\ref{fig:ngc1068_mspd_zoom}). Starting from the left hand side of the 2D wavelet space (i.e., shorter observed wavelength), for each ridge, we track the bin numbers and CWT coefficients from the largest scale present down to $s=2$. Note that a ridge starting from larger scale might diverge into two or more ridges at smaller scales due to the presence of noise. Therefore, we stop tracking the ridges at $s=2$. The ridge of a potential line should meet the following empirical criteria: 
\begin{enumerate}
    \item Local maxima should appear at larger scales $s\gtrsim 6$;
    \item At least 20\% of bin numbers should be identical so that the ridge is relatively straight;
    \item The maximum CWT coefficient should be at least 2.7$\sigma$ above the average. 
\end{enumerate}
The quoted numbers are used for the present sample analysis, which can certainly be fine tuned for a case study. The wavelength corresponds to the bin number that appears in most scales of the selected ridge is assigned to the potential line. This choice of wavelength is arbitrary but not critical because it will be determined more accurately via spectral fitting (Section~\ref{sect_fitting}).

\begin{figure*}
\centering
\includegraphics[width=\hsize, trim={2.0cm 1.0cm 2.0cm 1.0cm}, clip]{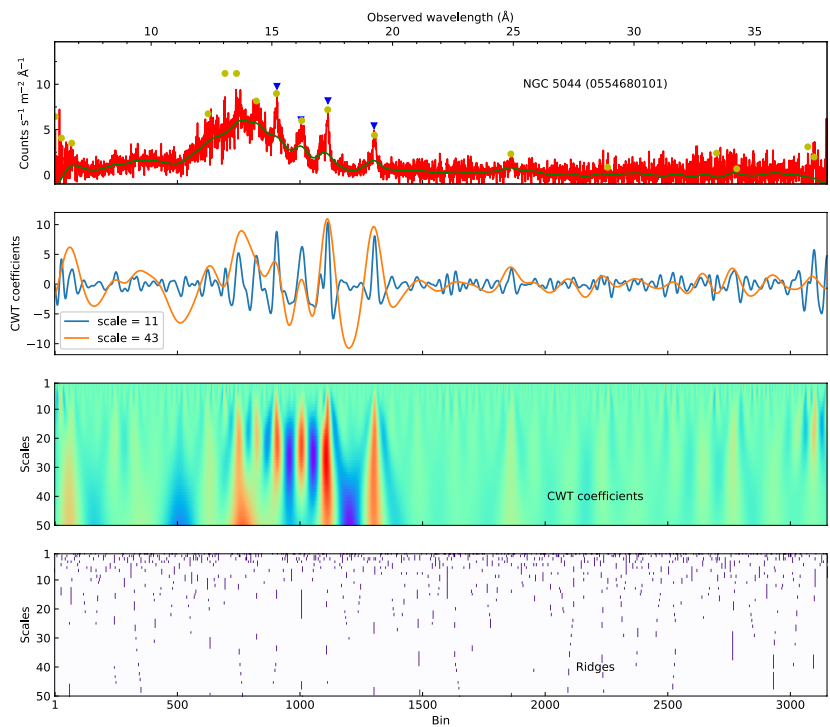}
\caption{Similar to Figure~\ref{fig:ngc1068_mspd} but for NGC\,5044 (ObsID: 0554680101).}
\label{fig:ngc5044_mspd}
\end{figure*}

While the present MSPD algorithm successfully detects the majority of visible peaks (by eye) in the point-like source NGC\,1068, we need to validate its performance on extended sources. Since the RGS spectrometer is slitless, for extended sources, their emission line profiles would appear broader, which complicates the analysis in practise. Figure~\ref{fig:ngc5044_mspd} shows the application of the present MSPD algorithm to the observed flux spectrum of NGC\,5044 (ObsID: 0554680101). Again, the majority of visible peaks (by eye) is detected. Meanwhile, the relatively poorer signal to noise ratio of the spectrum leads to more false detections. These false peaks will be filtered out after the following spectral fitting.

\subsection{Spectral fitting}
\label{sect_fitting}
The first-order background subtracted RGS~1 and RGS~2 spectra, as well as response matrices, are converted into the SPEX \citep{kaa96,kaa18} format through the SPEX task \textit{trafo}. For each observation, RGS~1 and RGS~2 spectra, if available, are fitted simultaneously. About $1~\%$ of the observations in the BIRD catalogue have either only RGS~1 or RGS~2 data, which are also fitted with data available. 

The full-width-half-maximum (${\rm FWHM}_{\rm inst}$) of the first-order RGS instruments is $\sim0.06-0.09~\AA$\footnote{We refer readers to Fig.~82 of the XMM User's hand book for details.}. For extended sources, due to the spatial broadening ($\theta \ge 0.8~{\rm arcmin}$ or 90\% of the point-spread-function) along the cross-dispersion direction, the effective ${\rm FWHM}_{\rm eff}$ is bigger than the instrumental ${\rm FWHM}_{\rm inst}$, 
\begin{equation}
    {\rm FWHM}_{\rm eff} = \frac{0.138}{m}\frac{\theta}{\rm arcmin} {\rm \AA},
\end{equation}
where $m$ is the spectral order \citep{tam04} and, in our case, $m=1$.

The default bin size of the RGS spectra is 0.01~\AA, which is clearly over-sampling. Therefore, we re-bin each RGS spectrum to achieve optimal binning \citep{kaa16}, which is $1/2-1/3$ of the maximum between ${\rm FWHM}_{\rm inst}$ and ${\rm FWHM}_{\rm eff}$. That is to say, in any case, the minimum binning factor is 3. 


Wavelength and flux values in CIELO are observed measurements. We did not apply any corrections for redshift or for X-ray absorption.

For each spectrum, the continuum is modelled with a spline function, which consists of a grid of nine points evenly distributed in the $6-38$~\AA\ wavelength range. The spline continuum modelling has been used successfully in e.g., \citet{det11}. 

Without any prior knowledge of the astrophysical properties of the targets, all emission features are modelled with Gaussian profiles. The multi-scale peak detection algorithm (Section~\ref{sect_mspd}) provides the number of emission features to SPEX, as well as the initial guess of the wavelength of each feature. The normalization, wavelength and broadening of each Gaussian are free to vary during the fit. Note that the velocity broadening of each Gaussian is constrained to be $0-10^4~{\rm km~s^{-1}}$. The upper limit is set to cover broad emission features from the broad-line region in active galactic nuclei \citep{bla90}.

The optimization method used here is the classical ``Levenberg-Marquardt" method, which might not lead to the global minimum. A better fit can be found during the error search of the parameters for each Gaussian. If so, we refit the spectrum, starting from those values that leads to better (or lower) statistics. We stop the iteration if the improvement on the $\Delta C \lesssim 0.5$, and limit of the number of iterations ($\le5$) if the convergence is slow. The slow convergence is partly due to those very broad emission features, where the broadening and the normalization parameters are degenerate to some extent. 

The best-fit parameters of an emission line is included in the final catalogue if its normalization is above zero at a $3\sigma$ confidence level and the energy flux is $\gtrsim10^{-18}~{\rm W~m^{-2}}$.

\begin{figure}
\footnotesize 
\centering
\includegraphics[width=0.9\hsize, trim={0.5cm 0.5cm 0.5cm 0.5cm}, clip]{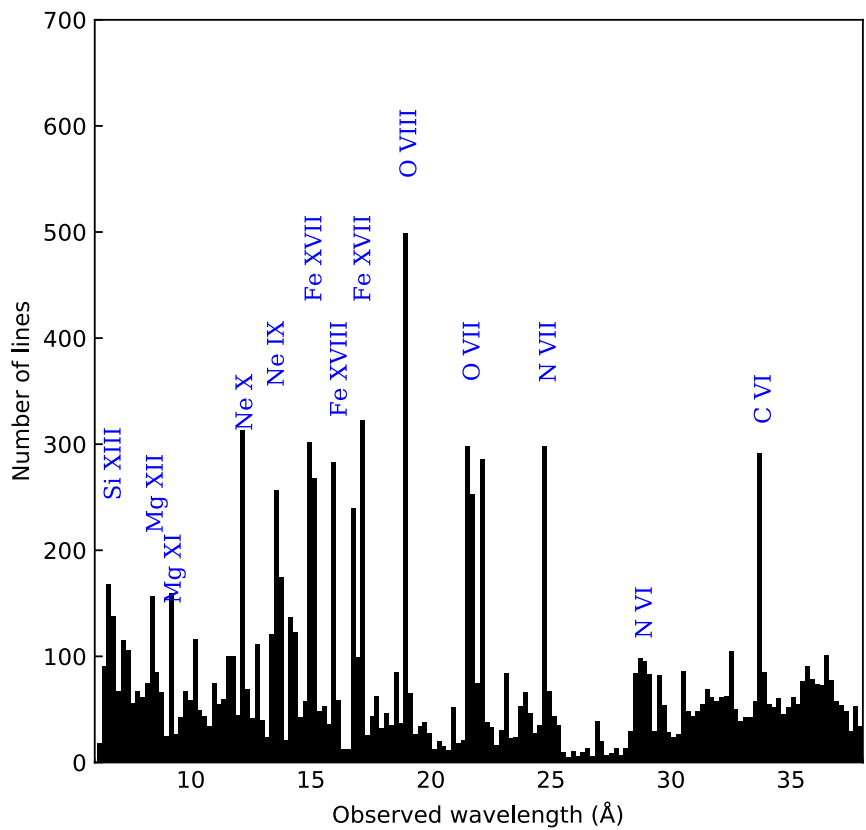}
\caption{The histogram of the CIELO catalogue as a function of observed wavelength. The rest frame wavelength of H-like Ly$\alpha$ and He-like resonances lines are labeled for guidance. }
\label{fig:hist_wav}
\end{figure}

\section{Catalogue structure}
\label{sect_structure}
The CIELO catalogue is in the form of machine readable ASCII table and FITS table. Both tables consist of 10981 rows from 1492 unique observations and 16 columns: Column $(1)$ is the ObsID; Columns $(2)-(4)$ are the wavelength (in \AA) and the $1\sigma$ lower and upper limits; Columns $(5)-(7)$ are the full width at half maximum (FWHM in \AA) and the $1\sigma$ lower and upper limits; Columns $(8)-(10)$ are normalization (in $10^{44}~{\rm ph~s^{-1}}$) and the $1\sigma$ lower and upper limits; Column $(11)$ is the observed energy flux (in ${\rm W~m^{-2}}$); Column $(12)$ is the observed photon flux (in ${\rm ph~m^{-2}~s^{-1}}$); Column $(13)$ is the modelled (spline) continuum photon flux density at the line center (in ${\rm ph~m^{-2}~s^{-1}~\AA^{-1}}$) and Column $(14)$ is the equivalent width (EW in $\AA$) defined as follows:
\begin{equation}
    EW = \frac{F_{\rm line}}{F_{\rm cont}}~,
\end{equation}
where $F_{\rm line}$ is the observed photon flux of the line (Col.~12) and $F_{\rm cont}$ is the modelled (spline) continuum photon flux density at the line center (Col.~13). The last two columns are the redshift (from Simbad, Col.~15) and target name (from XSA, Col.~16). 

\section{Results}
\label{sect_results}
CIELO includes data from $\sim66\%$ of the BIRD ''top-quality" observations. The targets of the remaining observations include some pulsars, BL-Lac objects, gamma-ray bursts, voids, etc. The soft X-ray spectra of these targets are featureless with small uncertainties on the (continuum) flux.

Figure~\ref{fig:hist_wav} gives an overview of the observed wavelength distribution of all the lines included in the catalog. H-like Lyman series, He-like triplets and Ne-like \ion{Fe}{XVII} resonance and forbidden lines dominates the catalog. Since observations are biased to nearby and bright targets, the observed wavelength of the emission lines peaks around the corresponding rest frame wavelength.

\subsection{Examples of CIELO results}
\label{sect_diagnostics}
Here we present comparisons of our results with published literature results. The comparisons are limited to point-like sources. As for extended sources like groups and clusters of galaxies \citep[e.g.,][]{kaa01,tam01,wer06} and supernovae remnants \citep[e.g.,][]{ras01,vhe03,bro11}, results of plasma modeling (e.g., temperature, abundance, emission measure) are usually provided, instead of a list of emission lines. 

\subsubsection{Local obscured AGN: NGC\,1068}
\begin{figure*}
\footnotesize 
\centering
\includegraphics[width=0.9\hsize, trim={0.cm 0.cm 0.cm 0cm}, clip]{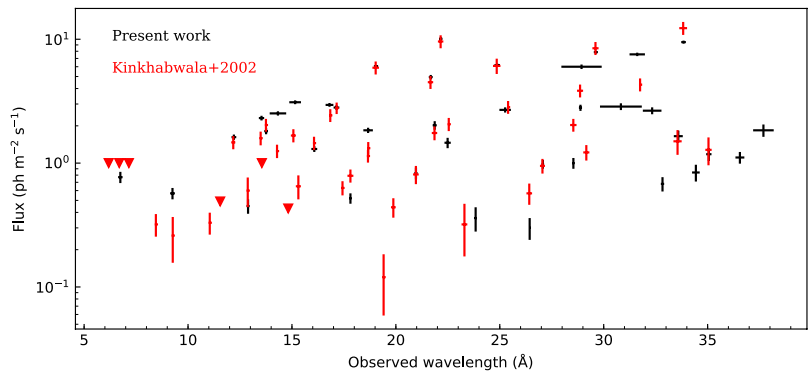}
\caption{Emission lines detected in the RGS spectrum of NGC\,1068 by the present work (black) and \citet[][red]{kin02}. For visibility, for each data point, we assign the line width (in $\AA$ instead of the wavelength uncertainty) to the horizontal error bar. }
\label{fig:cf_ngc1068}
\end{figure*}

NGC\,1068 is the prototypical Seyfert 2 galaxy in the local Universe with a redshift of $z=0.0038$ \citep{huc99}. Here we take the RGS spectrum of NGC\,1068 (ObsID: 0111200101, 41~ks effective exposure) as an example for local obscured AGN. We show in Figure~\ref{fig:cf_ngc1068} a comparison of the emission lines detected by the present work and \citet[][K02 hereafter]{kin02}. More details can be found in Table~\ref{tbl:ngc1068_line_list}. We also show in Figure~\ref{fig:0111200101_part_typ0} the results of spectral fitting (Section~\ref{sect_fitting}). 

K02 reported measurements of 40 lines and 6 RRC in their Table~1 and Table~2, respectively. The CIELO catalogue contains 38 lines (including RRC). The difference in the total number of lines is not unexpected. The present analysis has no prior knowledge of the atomic origin of the emission features. Consequently, for instance, the resonance and intercombination lines of \ion{Ne}{IX} are fitted as two lines in K02 yet as one line (i.e., a single Gaussian profile) here; the \ion{O}{VIII} Ly$\gamma$ and putative \ion{Fe}{XXIV}, \ion{Ar}{XI}, \ion{S}{XII} lines are missed in K02 but included here. Moreover, the present work is designed (Section~\ref{sect_mspd}) to omit relatively weak lines. Accordingly, weak lines like the \ion{O}{VIII} Ly$\delta$ line are included in K02 but missed here.

If lines are well separated, the velocity shifts ($v_{\rm off}$) and broadening ($\sigma_{\rm v}$), and the observed photon fluxes ($F_{\rm pho}$) from the above two analyses are consistent with each other. The \ion{Ne}{X} Ly$\alpha$ line is an exception, where the observed wavelength and velocity broadening are $(12.180\pm0.004)~\AA$ and $470\pm120~{\rm km~s^{-1}}$ in K02 yet $(12.206\pm0.005)~\AA$ and $1200\pm160~{\rm km~s^{-1}}$ in the present work. The observed \ion{Ne}{X} Ly$\alpha$ line profile shown in Figure~\ref{fig:0111200101_part_typ0} deviates to a Gaussian profile, which leads to the measurement bias. We also note that 12.165~\AA\ and $1400_{-14}^{+200}~{\rm km~s^{-1}}$ are quoted by \citet[][K14 hereafter]{kal14}, where the spectral analysis is based on the 450~ks \textit{Chandra} high-energy transmission grating spectroscopy data. 

If lines are blended with each other, disagreement on the velocity shift and broadening are expected. The \ion{Ne}{IX} He-like triplets around $13.5~\AA$ is a typical example. Note that both the present work and K14 cannot distinguish the resonance and intercombination lines. 

Furthermore, at shorter wavelength ($\lambda \lesssim 20~\AA$), the line widths obtained in the present work are in general larger than those given in K02, but within the upper limits given by K14. Better agreement is found at longer wavelength ($\lambda \gtrsim 20~\AA$).

There are two caveats in Table~\ref{tbl:ngc1068_line_list}. First, K02 assign a single value of 6.69~\AA\ to the rest frame wavelength of the \ion{Si}{XIII} He-like triplets. The wavelength of the resonance, intercombination, and forbidden lines are 6.648~\AA~(w), 6.685~\AA~(x), 6.688~\AA~(y), and 6.739~\AA~(z), respectively. Second, the present work fails to interpret the intercombination lines (x and y) of \ion{N}{VI} around 29~\AA~properly. The line profile has a single peak in RGS~1 but a double peak in RGS~2 (the fourth panel of Figure~\ref{fig:0111200101_part_typ0}). This leads to significantly biased results. K02 tied the velocity broadening of the He-like triplets and yielded more proper results. 

\subsubsection{Stars: Procyon and HD\,159176}
High quality RGS spectra of various active stars are available. We take Procyon and HD\,159176 as two examples to illustrate results from the present analysis.

\begin{figure*}
\footnotesize 
\centering
\includegraphics[width=0.9\hsize, trim={0.cm 0.cm 0.cm 0cm}, clip]{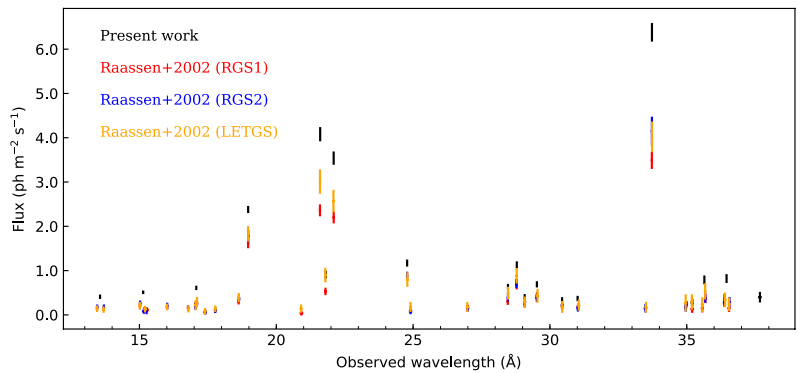}
\caption{Emission lines detected in the RGS spectrum of Procyon by the present work (black) and \citet[][red for RGS1, blue for RGS2, and orange for LETGS]{raa02}. The wavelength uncertainty is not visible here.}
\label{fig:cf_procyon}
\end{figure*}

The X-ray spectrum of Procyon is dominated by the late-type (F5 IV-V) optically bright star (with a faint white dwarf companion). The stellar corona has a temperature of $\sim(1-3)\times10^{6}~{\rm K}$ and exhibits a cooler X-ray spectrum than magnetically active stars \citep[][R02 hereafter]{raa02}. In Figure~\ref{fig:cf_procyon}, we compare the observed wavelengths and fluxes of ObsID 0123940101 (45.2~ks effective exposure) with those from Table~1 of R02, where both RGS and LEGS results are available. More details can be found in Table~\ref{tbl:procyon_line_list}. We also show in Figure~\ref{fig:0123940101_part_typ0} the results of spectral fitting (Section~\ref{sect_fitting}). 

Apart from those lines we fail to identify separately here (e.g., around 13.56~\AA\ and 36.45~\AA), a good agreement can be found between the present work and R02 in terms of the observed wavelength. As for the line (photon) fluxes, we find that larger values in the present work in addition to merely a few consistent results (e.g., the \ion{O}{VIII} 18.628 line). This is mainly due to the two different approaches for the continuum. As described in Section~\ref{sect_fitting}, a spline function is used here for the global continuum. On the other hand, R02 adopted a constant local ``background" level. The local background was adjusted for each emission line to account for the real continuum and the pseudo-continuum created by the overlap of several neglected weak lines. 

Unlike the narrow and ``regular" emission lines in Procyon, the emission lines in HD\,159176 are rather broad and asymmetric (neither Gaussian nor Lorentzian). HD\,159176 is a double-lined spectroscopic binary (O7V + O7V) in the young open cluster NGC\,6383 \citep{tru30}. As can be seen in Figure~\ref{fig:0001730201_part_typ0}, the Gaussian profiles adopted in the present analysis manage to capture the main features yet fail to account for the details. We also note that fitting RGS1 and RGS2 simultaneously rejects those features that appear in only one of the instrument (e.g., around $\sim10~\AA$) and those features that appears only in one instrument (e.g., the one around $\sim9.3~\AA$). 

According to \citet{dbe04}, the \ion{O}{VIII} Ly$\alpha$ line has an estimated FWHM of 2500~${\rm km~s^{-1}}$. Moreover, it is slightly blue-shifted $\sim300-600~{\rm km~s^{-1}}$, though the highest peak in the profile is found near the rest frame wavelength 18.967~\AA\ \citep{dbe04}. The present analysis yields an FWHM of $(0.130\pm0.014)~\AA$, i.e., $(2060\pm220)~{\rm km~s^{-1}}$ and an observed wavelength of $(18.975\pm0.008)~\AA$, i.e., blue-shifted by $(130\pm120)~{\rm km~s^{-1}}$. Given the complicated nature of this spectrum, our results are acceptable.

\begin{figure}
\footnotesize 
\centering
\includegraphics[width=\hsize, trim={0.5cm 0.5cm 0.5cm 0.5cm}, clip]{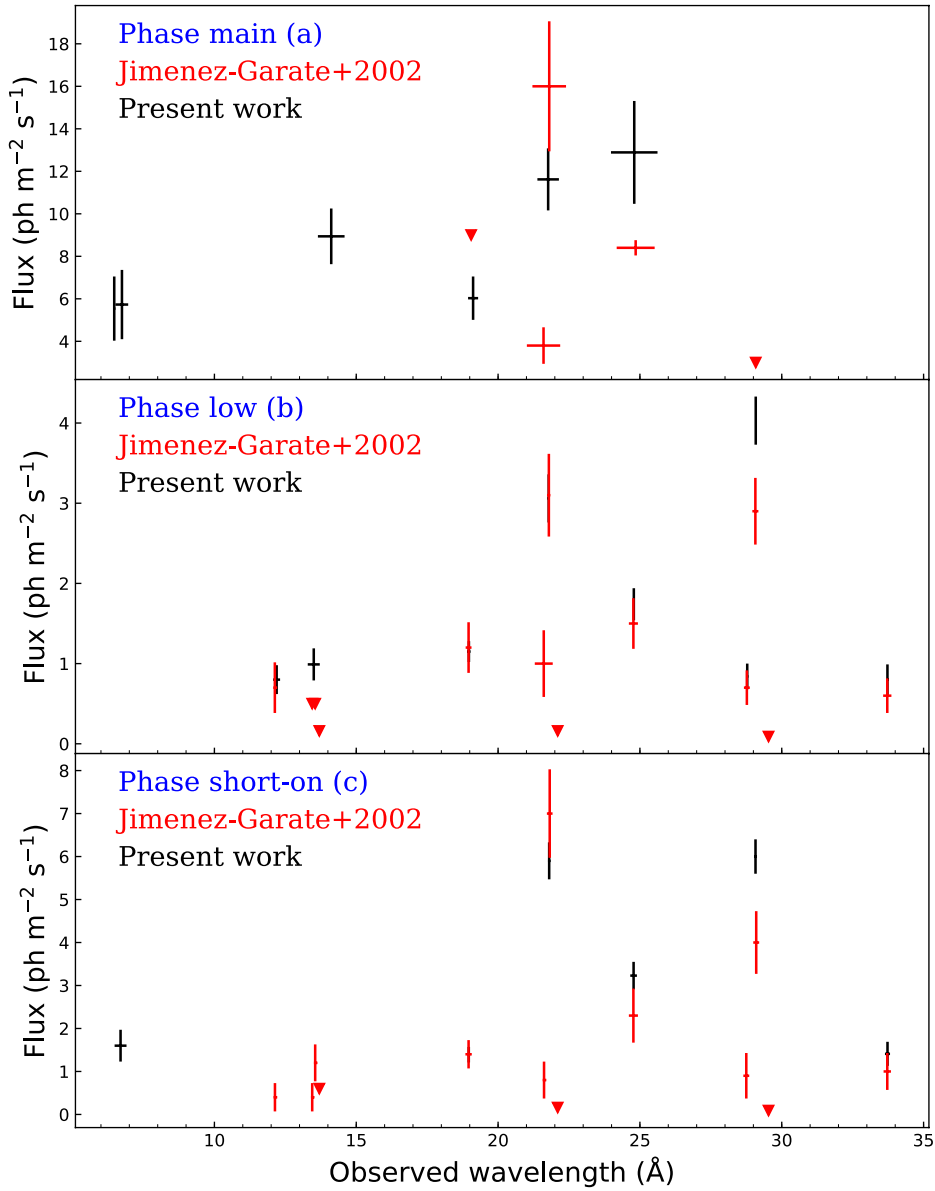}
\caption{A comparison of emission lines detected in the RGS spectra of Her X-1 in three different orbital phases. The present work and \citet{jig02} are shown in black and red, respectively.}
\label{fig:cf_her_x-1}
\end{figure}

\subsubsection{X-ray binaries: Her~X-1}
Her X-1 is a bright intermediate-mass X-ray binary \citep{sch72}. It exhibits an unusual long-term X-ray flux modulation with a period of 35 days \citep{gia73}. The 35-day X-ray light curve is asymmetric and contains two peaks: a state of 8-day duration reaching the peak flux $F_{\rm max}$ named the main-on state and a 4-day duration reaching one third of the maximum named the short-on state. The rest is named the low state with flux twenty times lower than the maximum \citep[][JG02]{jig02}. 

We inspect the RGS spectra of Her X-1 at three different orbital phases: 0134120101 (main), 0111060101 (low), and 0111061201 (short-on). Figure~\ref{fig:cf_her_x-1} visualizes the comparison of emission lines reported in the present work and JG02. Tabulated results are given in Table~\ref{tbl:herx1_line_list}. The plot of the spectral fitting for the short-on phase can be found in Figure~\ref{fig:0111061201_part_typ0}.

Minor differences are found between the present analysis and JG02. The present work does not include the weak resonance line in the \ion{O}{VII} He-like triplets. The \ion{N}{VII} Ly$\alpha$ line is found to be rather broad $(4200\pm1600)~{\rm km~s^{-1}}$ in the present work. The width was fixed to $3200~\rm{km~s^{-1}}$ in JG02, based on their width of \ion{O}{VII} He-like triplets. As a result, the line (photon) flux reported here is slightly higher (at a $2\sigma$ confidence level) than that of JG02.

Moreover, we also include a few broad features at $6.4~\AA$, $6.7~\AA$, and $14.1~\AA$. Such broad features are likely due to calibration uncertainties of the instrument. The smaller the statistical uncertainty, the more prominent such systematic uncertainties.  

\subsubsection{Comparison of plasma diagnostics: stars vs. obscured AGN}

The intensity of the brightest components of $n=2$ to $n=1$ transitions in He-like ions can be combined into ratios with a strong diagnostic power of the electron density [$R(n_e)$] and electronic temperature [$G(T_e)$] \citep{gabriel69}:
$$
R(n_e) \equiv \frac{f}{i}
$$
$$
G(T_e) \equiv \frac{f+i}{r}
$$
where $f$, $i$, and $r$ are the intensities of the forbidden ($z$; 1s$^2$~$^1$S$_0$ - 1s2s~$^3$S$_1$), intercombination ($x$,$y$; 1s$^2$~$^1$S$_0$ - 1s2p~$^3$P$_{2,1}$), and resonance ($w$; 1s$^2$~$^1$S$_0$ - 1s2p~$^1$P$_1$) lines, respectively. The same diagnostic ratios can be used to study photoionised plasma as well \citep{porter07}. Indeed, they can be used to identify the dominant ionization mechanism \citep{porquet00}.

\begin{figure}
\footnotesize 
\hspace{-1.5cm}
\includegraphics[width=0.8\hsize, angle=90, clip]{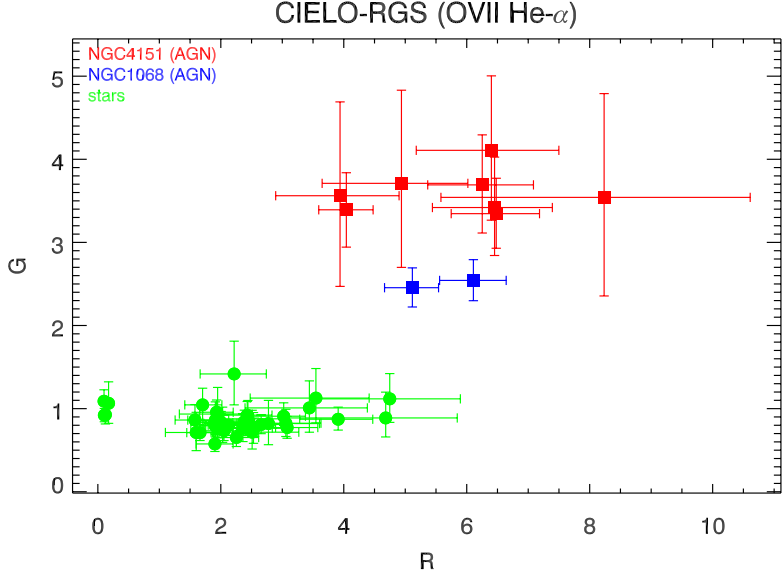}
\caption{OVII He-$\alpha$ diagnostic parameters $R$ and $G$ for the two brightest heavily obscured AGN (NGC~1068, \textit{blue}; NGC~4151, \textit{red}), and for a sample of active stars (\textit{green}) in CIELO-RGS.}
\label{fig:mg1}
\end{figure}

The CIELO catalogue allows to perform studies of these diagnostic parameters on large source populations. As a token of example, in Fig.~\ref{fig:mg1}
we show the $R$ vs. $G$ diagnostic plot for the OVII He-$\alpha$ transition derived from the CIELO catalogue for
the two soft X-ray brightest heavily obscured AGN (NGC~1068, and NGC~4151) \citep{gua07}, and for a sample of active stars. Multiple data points for each source correspond to different observations. The AGN X-ray spectra
are dominated by plasma photoionized by the intense radiation field of the active nucleus \citep{kin02}, most likely due to X-ray diffuse emission associated with the Extended Narrow Line Regions \citep{bianchi06}. The variability of the $R$ parameter reflects time changes in the ratio between the forbidden and the intercombination component that could be related to the UV illumination of the ionized nebula in a radiation-driven wind \citep{landt15}.
Stellar spectra should be instead dominated by optically thin collisionally ionized plasma. The CIELO diagnostic diagram nicely separates the two classes, as expected.

\section{Summary}
\label{sect_discussion}

In this paper we present CIELO-RGS, the first systematic catalogue of emission lines detected by the RGS instrument \citep{denherder01} on-board \textit{XMM-Newton}. CIELO-RGS was build using high-quality spectra accumulated during more then 5000 observations performed over 15~years of \textit{XMM-Newton} science operations. This catalogue is intended to be an additional tool to facilitate the exploitation of the RGS spectra available in the public science archive, in particular to perform correlation between X-ray spectral diagnostics parameters and measurements at other wavelengths.

The catalogue generation is based on a two-tier process. Firstly, each RGS spectrum was analyzed searching for candidate emission lines with a technique fully independent of the spectral modeling, and therefore in principle of the intrinsic physical and astrophysical nature of the emitting plasma. This technique is a version of a multi-scale peak detection algorithm, robustly tested in other scientific contexts (Z15). This step provides a list of candidate emission lines detected in each individual observation spectrum over the appropriate range of frequency scales (upper-bound by the instrumental resolution). This list of candidate emission lines is subsequently validated through a formal forward-folding spectral fitting process that takes properly into account the instrumental transfer function, the observation background, and the rigorous statistical conditions for spectral binning \citep{kaa16}. The resulting catalogue include the following measured observable for each line: wavelength, intrinsic profile width, normalization, energy and photon flux, and equivalent width, as well as a phenomenological parameterization of the underlying continuum.

In this paper we present some illustrative examples of the wide range of scientific themes than such a catalogue can address, from the ionization mechanism and dynamics of the Extended Narrow Line Region in AGN, to plasma diagnostics in optically thin, collisionally ionised plasma in stars or the intra-cluster medium permeating galaxy clusters, to coronal emission in accreting stellar-mass black holes. These topics will be further developed in future specific papers. 

\begin{acknowledgements}
We thank the referee for the careful reading of the manuscript. The authors gratefully acknowledge a generous financial support by the Divisional Funding of the Science Support Office at ESA, in the framework of the EXPRO contract 40000124950/18/NL/JB/gg. J. M. acknowledges discussions and consultations with J. de Plaa, N. Loiseau, M. Mehdipour, and C. de Vries. SRON is supported financially by NWO, the Netherlands Organization for Scientific Research.
\end{acknowledgements}


\appendix
\section{Detailed comparisons of the examples}
\begin{figure*}
\footnotesize 
\centering
\includegraphics[width=\hsize, trim={1.cm 1.cm 2.cm 1cm}, clip]{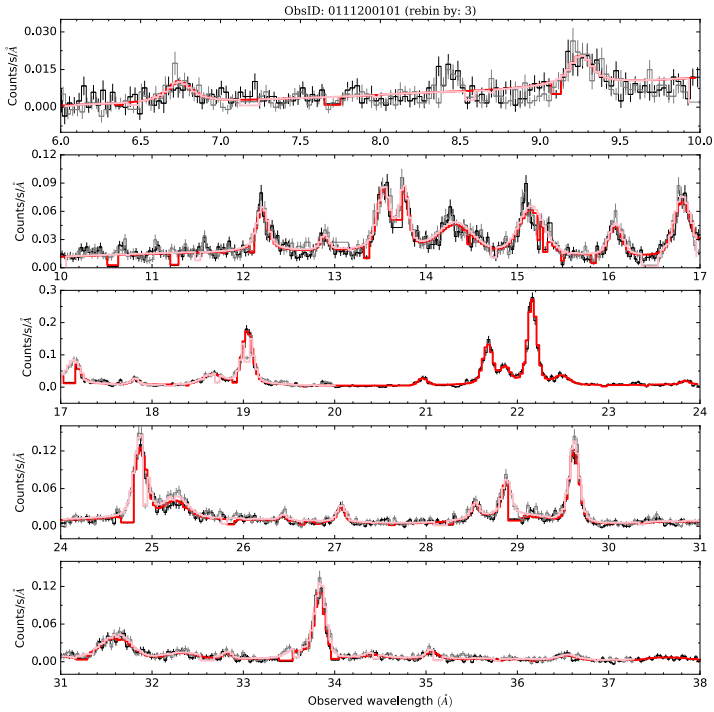}
\caption{The RGS spectrum of NGC\,1068 (ObsID: 0111200101). RGS~1 and RGS~2 spectra are fitted simultaneously. RGS spectra are rebinned by a factor of 3. RGS~1 and RGS~2 data are shown in black and gray, respectively. The best-fit models are shown in red and pink for RGS~1 and RGS~2, respectively. The breaks (e.g., around 11.2~\AA\ and 24~\AA) are due to CCD gaps, bad pixels, etc.}
\label{fig:0111200101_part_typ0}
\end{figure*}

\begin{figure*}
\footnotesize 
\centering
\includegraphics[width=\hsize, trim={1.cm 1.cm 2.cm 1cm}, clip]{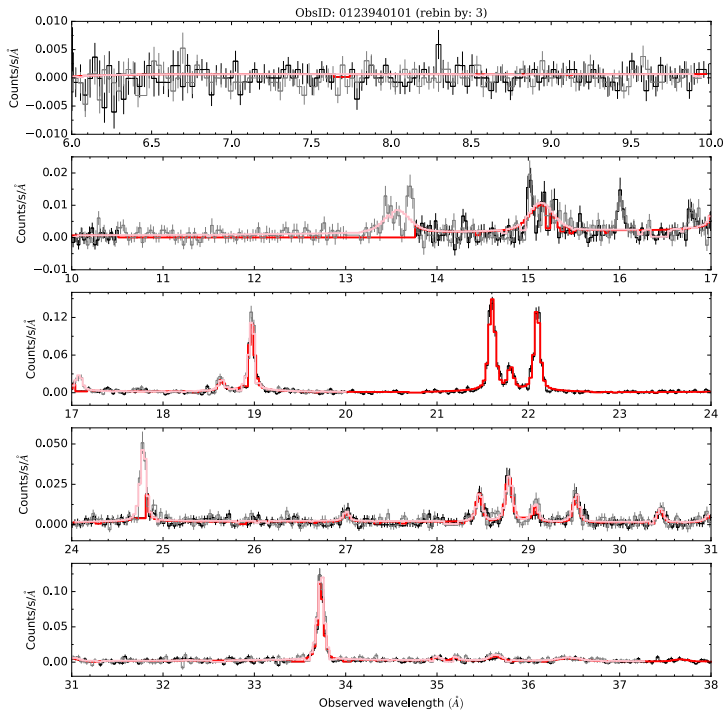}
\caption{Similar to Figure~\ref{fig:0123940101_part_typ0} but for Procyon (ObsID: 0123940101).}
\label{fig:0123940101_part_typ0}
\end{figure*}

\begin{figure*}
\footnotesize 
\centering
\includegraphics[width=\hsize, trim={1.cm 1.cm 2.cm 1cm}, clip]{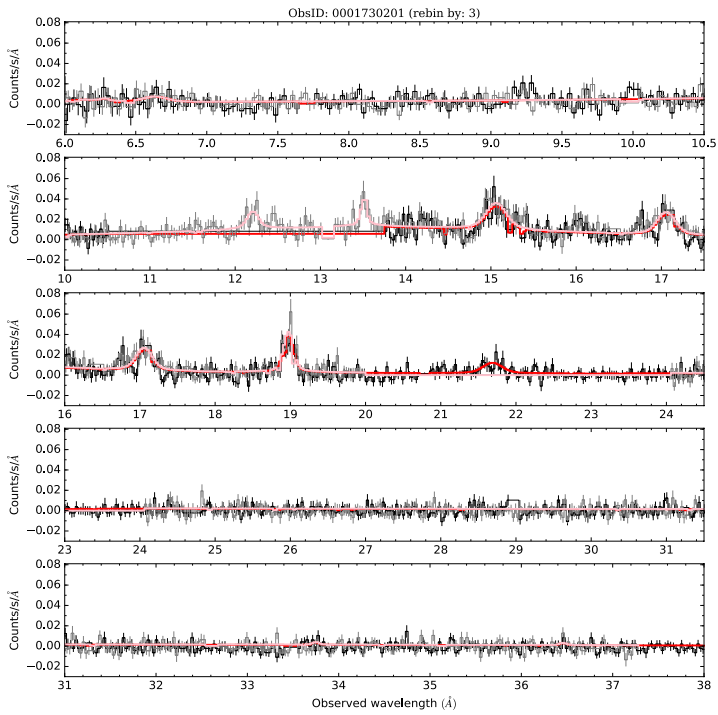}
\caption{Similar to Figure~\ref{fig:0111200101_part_typ0} but for HD\,159176 (ObsID: 0001730201).}
\label{fig:0001730201_part_typ0}
\end{figure*}

\begin{figure*}
\footnotesize 
\centering
\includegraphics[width=\hsize, trim={1.cm 1.cm 2.cm 1cm}, clip]{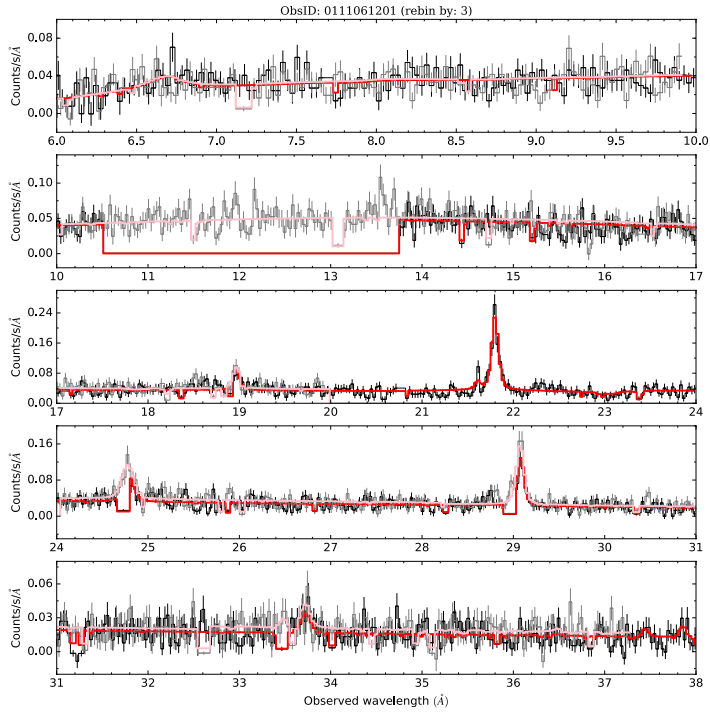}
\caption{Similar to Figure~\ref{fig:0111200101_part_typ0} but for Her X-1 (ObsID: 0111061201).}
\label{fig:0111061201_part_typ0}
\end{figure*}

We show the best-fit models against the observed spectra for NGC\,1068, Procyon, HD\,15176, and Her X-1 in Section~\ref{sect_results} (Figure~\ref{fig:0111200101_part_typ0} to \ref{fig:0111061201_part_typ0}). We also tabulated comparisons of the emission lines among the present work and others (Table~\ref{tbl:ngc1068_line_list} to \ref{tbl:herx1_line_list}). 

\begin{table*}
\caption{List of lines and radiative recombination continua (RRC) in NGC\,1068. Measurements on the left-hand-side (Cols.~$3-5$) are from \citet{kin02}, while those on the right-hand-side (Cols.~$6-9$) are from the present work.}
\label{tbl:ngc1068_line_list}
\centering
\begin{tabular}{lc|ccc|ccccccc}
\hline\hline
\noalign{\smallskip} 
Line ID & $\lambda_0$ & $v_{\rm off}$ & $\sigma_v$ & $F_{\rm pho}$ & $\lambda_{\rm obs}$ & $v_{\rm off}$ & $\sigma_v$ & $F_{\rm pho}$ \\
 & $\AA$ & ${\rm km~s^{-1}}$ & ${\rm km~s^{-1}}$ & ${\rm ph~m^{-2}~s^{-1}}$ & $\AA$ & ${\rm km~s^{-1}}$ & ${\rm km~s^{-1}}$ & ${\rm ph~m^{-2}~s^{-1}}$ \\
\noalign{\smallskip} 
\hline
\noalign{\smallskip} 
\ion{Si}{XIV} Ly$\alpha$ & 6.182 & ... & ... & $\lesssim0.8\pm0.2$ & \\
\ion{Si}{XIII} & 6.69 & ... & ... & $\lesssim0.8\pm0.2$ & 6.74$\dagger$ & ... & ... & $0.77\pm0.08$ \\
\ion{Si}{I} K & 7.154 & ... & ... & $\lesssim0.8\pm0.2$ & \\
\ion{Mg}{XII} Ly$\alpha$ & 8.421 & $70\pm250$ & ... & $0.32\pm0.06$ & \\
\ion{Mg}{XI} & 9.23 & ... & ... & $0.26\pm0.10$ & 9.252 & $-430\pm390$ & $1600\pm360$ & $0.57\pm0.06$ \\
\ion{Ne}{IX} He$\gamma$ & 11.000 & $110\pm220$ & ... & $0.33\pm0.06$ \\ 
\ion{Ne}{IX} He$\beta$ & 11.547 & ... & ... & $\lesssim0.42\pm0.07$ \\ 
\ion{Ne}{X} Ly$\alpha$ & 12.134 & $-100\pm200$ & $470\pm120$ & $1.47\pm0.15$ & 12.206 & $640\pm120$ & $1200\pm160$ & $1.62\pm0.08$ \\
\ion{Fe}{XX} & 12.846 & $-500\pm190$ & ... & $0.60\pm0.15$ & 12.888 & $-160\pm210$ & $710\pm240$ & $0.45\pm0.06$ \\
\ion{Ne}{IX} $r$ & 13.447 & $-90\pm180$ & ... & $1.59\pm0.17$ & \multirow{2}{*}{13.535$\dagger$} & \multirow{2}{*}{...} & \multirow{2}{*}{...} & \multirow{2}{*}{$2.31\pm0.10$} \\ 
\ion{Ne}{IX} $i$ & 13.552 & ... & ... & $\lesssim0.83\pm0.17$ \\
\ion{Ne}{IX} $f$ & 13.698 & $-30\pm180$ & ... & $2.03\pm0.20$ & 13.756 & $130\pm90$ & $830\pm110$ & $1.81\pm0.10$ \\
\ion{O}{VIII} RRC & 14.228 & ... & ... & $1.25\pm0.13$ & \multirow{2}{*}{14.326$\dagger$} & \multirow{2}{*}{...} & \multirow{2}{*}{...} & \multirow{2}{*}{$2.52\pm0.10$} \\
\ion{O}{VIII} Ly$\delta$ & 14.821 & ... & ... & $\lesssim0.37\pm0.06$ & \\
\ion{Fe}{XVII} & 15.014 & $-60\pm160$ & $570\pm100$ & $1.67\pm0.17$ & \multirow{3}{*}{15.153$\dagger$} & \multirow{3}{*}{...} & \multirow{3}{*}{...} & \multirow{3}{*}{$3.10\pm0.11$} \\
\ion{O}{VIII} Ly$\gamma$ & 15.176 & ... & ... & ... & \\
\ion{Fe}{XVII} & 15.261 & $-210\pm160$ & $570\pm100$ & $0.65\pm0.13$ & \\
\ion{O}{VIII} Ly$\beta$ & 16.006 & $-90\pm150$ & ... & $1.45\pm0.15$ & 16.075 & $150\pm110$ & $1100\pm110$ & $1.30\pm0.07$ \\
\ion{O}{VII} RRC & 16.78 & ... & ... & $2.43\pm0.24$ & 16.806 & ... & ... & $2.95\pm0.10$ \\
\ion{Fe}{XVII} & 17.051, 17.096 & $-180\pm140$ & ... & $2.78\pm0.28$ & 17.142$\dagger$ & ... & ... & $2.81\pm0.12$ \\
\ion{O}{VII} He$\delta$ & 17.396 & $-520\pm140$ & ... & $0.63\pm0.07$ \\
\ion{O}{VII} He$\gamma$ & 17.768 & $-380\pm140$ & $640\pm80$ & $0.79\pm0.08$ & 17.814 & $-360\pm100$ & $350\pm220$ & $0.52\pm0.05$ \\ 
\ion{N}{VII} RRC & 18.587 & ... & ... & $1.14\pm0.11$ & \\
\ion{O}{VII} He$\beta$ & 18.627 & $-430\pm130$ & ... & $1.32\pm0.13$ & 18.664 & $-540\pm130$ & $1570\pm140$ & $1.84\pm0.09$ \\ 
\ion{O}{VIII} Ly$\alpha$ & 18.969 & $-200\pm130$ & $700\pm80$ & $5.89\pm0.59$ & 19.044 & $50\pm30$ & $690\pm30$ & $6.06\pm0.14$ \\
\ion{N}{VII} Ly$\delta$ & 19.361 & $-380\pm120$ & ... & $0.12\pm0.06$ & \\
\ion{N}{VII} Ly$\gamma$ & 19.826 & $-350\pm120$ & $390\pm80$ & $0.44\pm0.07$ & \\
\ion{N}{VII} Ly$\beta$ & 20.910 & $-370\pm120$ & $520\pm70$ & $0.81\pm0.12$ & 20.967 & $-320\pm100$ & $450\pm150$ & $0.83\pm0.09$ \\
\ion{O}{VII} $r$ & 21.602 & $-260\pm110$ & $450\pm70$ & $4.50\pm0.45$ & 21.678 & $-90\pm40$ & $550\pm50$ & $4.95\pm0.20$ \\ 
\ion{O}{VII} $i$ & 21.803 & $-380\pm110$ & $450\pm70$ & $1.75\pm0.19$ & 21.869 & $-230\pm100$ & $570\pm120$ & $2.02\pm0.16$ \\
\ion{O}{VII} $f$ & 22.101 & $-430\pm110$ & $450\pm70$ & $9.59\pm0.96$ & 22.164 & $-290\pm30$ & $460\pm40$ & $10.08\pm0.28$ \\
\ion{N}{VI} RRC & 22.46 & ... & ... & $2.06\pm0.21$ & 22.490 & ... & ... & $1.46\pm0.14$ \\
\ion{N}{VI} He$\delta$ & 23.277 & $-210\pm100$ & $470\pm60$ & $0.32\pm0.14$ \\
\ion{N}{VI} He$\gamma$ & 23.771 & $-490\pm100$ & $360\pm60$ & $0.77\pm0.12$ & 23.838 & $-290\pm190$ & $300\pm200$ & $0.36\pm0.08$ \\ 
\ion{N}{VII} Ly$\alpha$ & 24.781 & $-270\pm100$ & $650\pm60$ & $6.09\pm0.74$ & 24.876 & $10\pm40$ & $690\pm40$ & $6.18\pm0.18$ \\
\ion{C}{VI} RRC & 25.303 & ... & ... & $2.83\pm0.28$ & 25.258 & ... & ... & $2.69\pm0.13$ \\
\ion{C}{VI} Ly$\delta$ & 26.357 & $-440\pm90$ & $400\pm60$ & $0.57\pm0.10$ & 26.438 & $-220\pm130$ & $<140$ & $0.30\pm0.06$ \\
\ion{C}{VI} Ly$\gamma$ & 26.990 & $-440\pm90$ & $320\pm60$ & $0.95\pm0.11$ & 27.073 & $-220\pm60$ & $280\pm80$ & $0.98\pm0.08$ \\
\ion{C}{VI} Ly$\beta$ & 28.466 & $-380\pm80$ & $450\pm50$ & $2.03\pm0.20$ & 28.541 & $-350\pm60$ &  $230\pm120$ & $1.00\pm0.10$ \\
\ion{N}{VI} $r$ & 28.787 & $-350\pm80$ & $410\pm50$ & $3.84\pm0.38$ & 28.882 & $-150\pm40$ & $330\pm50$ & $2.80\pm0.16$ \\ 
\ion{N}{VI} $i$ & 29.083 & $-310\pm80$ & $410\pm50$ & $1.22\pm0.15$ & 28.933$\dagger$ & ... & ... & $6.00\pm0.25$ \\
\ion{N}{VI} $f$ & 29.534 & $-430\pm80$ & $410\pm50$ & $8.46\pm0.85$ & 29.622 & $-250\pm20$ & $420\pm30$ & $7.87\pm0.22$ \\
\ion{Fe}{XXIV} $\triangledown$ & 30.746 & ... & ... & ... & 30.827 & ... & ... & $2.86\pm0.18$ \\
\ion{C}{V} RRC & 31.63 & ... & ... & $4.30\pm0.43$ & 31.613 & ... & ... & $7.55\pm0.24$ \\
\ion{S}{XIII} $\triangledown$ & 32.239 & ... & ... & ... & 32.331 & ... & ... & $2.65\pm0.17$ \\
\ion{C}{V} He$\delta$ $\triangledown$ & 32.754 & ... & ... & ... & 32.829 & ... & ... & $0.68\pm0.09$ \\
\ion{C}{V} He$\gamma$ & 33.426 & ... & $550\pm50$  & $1.50\pm0.31$ & 33.589 & $320\pm140$ & $820\pm100$ & $1.65\pm0.18$ \\ 
\ion{C}{VI} Ly$\alpha$ & 33.736 & $-360\pm70$ & $510\pm40$ & $12.29\pm1.23$ & 33.842 & $-200\pm20$ & $420\pm20$ & $9.48\pm0.27$ \\
\ion{Ar}{XI} $\triangledown$  & 34.330 & ... & ... & ... & 34.435 & ... & ... & $0.84\pm0.13$ \\
\ion{C}{V} He$\beta$ & 34.973 & $-550\pm70$ & $360\pm40$ & $1.28\pm0.30$ & 35.058 & $-410\pm90$ & $470\pm80$ & $1.18\pm0.14$ \\
\ion{S}{XII} $\triangledown$ & 36.398 & ... & ... & ... & 36.553 & ... & ... & $1.11\pm0.12$ \\
\ion{Ca}{XII} $\triangledown$ & 37.604 & ... & ... & ... & 37.685 & ... & ... & $1.84\pm0.21$ \\
\noalign{\smallskip} 
\hline
\end{tabular}
\tablefoot{$v_{\rm off}$ is the velocity shift. $\sigma_v$ is the velocity broadening. $F_{\rm pho}$ is the observed photon flux. Those line IDs labeled with a down triangle ($\triangledown$) are inconclusive. Those labeled with dagger ($\dagger$) correspond to multiple lines. }
\end{table*}

\begin{table*}
\caption{List of lines in Procyon. Measurements on the left-hand-side (Cols.~$3-8$) are from \citet{raa02}, while those on the right-hand-side (Cols.~9 and 10) are from the present work.}
\label{tbl:procyon_line_list}
\centering
\small
\begin{tabular}{lc|cccccc|cc}
\hline\hline
\noalign{\smallskip} 
Line ID & $\lambda_0$ & $\lambda_{\rm R1}$ & $F_{\rm R1}$ & $\lambda_{\rm R2}$ & $F_{\rm R2}$ & $\lambda_{\rm L}$ & $F_{\rm L}$ & $\lambda_{\rm R}$ & $F_{\rm R}$   \\
 & $\AA$ & $\AA$ & ${\rm ph~m^{-2}~s^{-1}}$ & $\AA$ & ${\rm ph~m^{-2}~s^{-1}}$ & $\AA$ & ${\rm ph~m^{-2}~s^{-1}}$ & $\AA$ & ${\rm ph~m^{-2}~s^{-1}}$  \\
\noalign{\smallskip} 
\hline
\noalign{\smallskip} 
\ion{Ne}{IX} $r$ & 13.448 & ... & ... & 13.454(9) & 0.17(4) & 13.450(13) & 0.14(5) & \multirow{2}{*}{$13.561\pm0.017\dagger$} & \multirow{2}{*}{$0.41\pm0.05$} \\
\ion{Ne}{IX} $r$ & 13.700 & ... & ... & 13.701(6) & 0.17(4) & 13.690(15) & 0.12(5) & \\
\ion{Fe}{XVII}$\dagger$ & 15.014 & 15.018(7) & 0.20(5) & 15.025(9) & 0.26(4) & 15.015(18) & 0.21(6) & \multirow{3}{*}{$15.138\pm0.016\dagger$} & \multirow{3}{*}{$0.51\pm0.04$} \\
\ion{O}{VIII} & 15.175 & 15.176(13) & 0.12(3) & 15.161(21) & 0.08(3) & ... & ... & \\
\ion{Fe}{XVII} & 15.265 & 15.281(12) & 0.11(3) & 15.258(12) & 0.09(5) & ... & ... &  \\
\ion{Fe}{XVII} & 16.780 & 16.776(9) & 0.16(3) & 16.776(15) & 0.12(4) & 16.790(14) & 0.13(5) &  \\
\ion{Fe}{XVII} & 17.055 & 17.047(9) & 0.23(4) & 17.044(14) & 0.19(5) & 17.054(12) & 0.22(9) & \multirow{2}{*}{$17.077\pm0.005\dagger$} & \multirow{2}{*}{$0.61\pm0.05$} \\
\ion{Fe}{XVII} & 17.100 & ... & ... & 17.099(12) & 0.25(5) & 17.102(9) & 0.29(9) & \\
\ion{O}{VIII} & 17.380 & 17.402(12) & 0.07(3) & 17.402(28) & 0.06(3) & 17.396(17) & 0.08(5) & \\
\ion{O}{VIII} & 17.770 & 17.765(9) & 0.10(3) & 17.772(9) & 0.11(4) & 17.796(13) & 0.14(5) &  \\
\ion{O}{VIII} & 18.628 & 18.624(4) & 0.30(4) & 18.637(6) & 0.37(5) & 18.629(5) & 0.39(8) & $18.638\pm0.006$ & $0.35\pm0.04$ \\
\ion{O}{VIII} & 18.973 & 18.970(3) & 1.61(8) & 18.975(2) & 1.78(11) & 18.972(2) & 1.83(15) & $18.972\pm0.006$ & $2.38\pm0.08$ \\
\ion{N}{VII} & 20.910 & 20.198(27) & 0.04(3) & ... & ... & 20.905(22) & 0.14(7) & \\
\ion{O}{VII} & 21.602 & 21.596(2) & 2.36(11) & ... & ... & 21.597(2) & 3.01(25) & $21.604\pm0.002$ & $4.08 \pm0.16$ \\
\ion{O}{VII} & 21.800 & 21.797(4) & 0.53(6) & ... & ... & 21.792(5) & 0.90(14) & $21.807\pm0.005$ & $0.91\pm0.09$ \\
\ion{O}{VII} & 22.100 & 22.098(2) & 2.20(11) & ... & ... & 22.089(2) & 2.57(23) & $22.097\pm0.004$ & $3.54\pm0.15$ \\
\ion{N}{VII} & 24.781 & ... & ... & 24.780(3) & 0.88(7) & 24.790(4) & 0.80(14) & $24.782\pm0.006$ & $1.17\pm0.08$ \\
\ion{N}{VI} & 24.900 & ... & ... & 24.907(22) & 0.08(4) & 24.906(11) & 0.18(9) & \\
\ion{C}{VI} & 26.990 & 27.001(11) & 0.16(4) & 26.994(12) & 0.14(4) & 26.979(19) & 0.18(9) & $27.003\pm0.010$ & $0.17\pm0.03$ \\
\ion{C}{VI} & 28.470 & 28.460(7) & 0.30(5) & 28.465(6) & 0.39(6) & 28.470(6) & 0.49(12) & $28.466\pm0.005$ &  $0.64\pm0.06$ \\
\ion{N}{VI} & 28.790 & 28.785(4) & 0.69(9) & 28.775(6) & 0.71(8) & 28.785(5) & 0.88(15) & $28.789\pm0.003$  & $1.13\pm0.08$ \\
\ion{N}{VI} & 29.090 & 29.078(9) & 0.24(5) & 29.084(9) & 0.29(7) & 29.082(12) & 0.29(10) & $29.077\pm0.009$ &  $0.41\pm0.06$ \\
\ion{N}{VI} & 29.530 & 29.524(6) & 0.43(6) & 29.520(8) & 0.38(6) & 29.546(11) & 0.43(13) & $29.523\pm0.008$ & $0.69\pm0.07$ \\
\ion{Ca}{XI} & 30.448 & 30.445(12) & 0.20(5) & 30.446(13) & 0.22(5) & 30.450(25) & 0.18(11) & $30.444\pm0.009$ & $0.35\pm0.05$ \\
\ion{Si}{XII} & 31.015 & 31.027(16) & 0.18(5) & 31.021(12) & 0.15(5) & 31.054(15) & 0.22(10) & $31.011\pm0.014$ &  $0.37\pm0.06$ \\
\ion{S}{XIV}$\triangledown$ & 33.549 & ... & ... & 33.490(26) & 0.15(7) & 33.510(18) & 0.17(10) &  \\
\ion{C}{VI} & 33.736 & 33.724(2) & 3.49(17) & 33.726(2) & 4.15(30) & 33.731(2) & 4.02(32) & $33.728\pm0.001$ & $6.38\pm0.21$ \\
\ion{C}{V} & 34.970 & 34.967(12) & 0.19(6) & 34.962(15) & 0.17(7) & 34.959(15) & 0.27(17) & $34.994\pm0.021$  & $0.25\pm0.06$ \\
\ion{Ca}{XI} & 35.212 & 35.198(30) & 0.13(6) & 35.193(12) & 0.27(8) & 35.188(18) & 0.29(15) & $35.202\pm0.020$ & $0.29\pm0.07$ \\
\ion{Ca}{XI} & 35.576 & 35.566(16) & 0.15(7) & 35.562(12) & 0.26(7) & 35.566(16) & 0.23(14) &  \\
\ion{S}{XIII} & 35.665 & 35.682(8) & 0.37(8) & 35.676(9) & 0.38(8) & 35.672(9) & 0.53(16) & $35.644\pm0.010$ & $0.79\pm0.10$ \\
\ion{S}{XII} & 36.398 & 36.374(12) & 0.35(10) & 36.372(15) & 0.28(7) & 36.399(15) & 0.34(14) & $36.454\pm0.018\dagger$ & $0.82\pm0.10$ \\
\ion{S}{XII} & 36.563 & 36.544(19) & 0.15(6) & 36.561(19) & 0.29(9) & 36.547(15) & 0.24(13) &  \\
\ion{S}{XIII}$\triangledown$ & 36.72 & ... & ... & ... & ... & ... & ... & $37.674\pm0.065\dagger$ & $0.40\pm0.12$ \\
\noalign{\smallskip} 
\hline
\end{tabular}
\tablefoot{Those line IDs labeled with a down triangle ($\triangledown$) are inconclusive. Those labeled with dagger ($\dagger$) correspond to multiple lines. }
\end{table*}

\begin{table*}
\caption{List of lines for Her X-1 at different orbital phases. The main-on (orbital phase $0.18-0.26$), low ($0.47-0.56$), and short-on ($0.52-0.60$) states are noted as a (0134120101), b (0111060101), and c (0111061201). Measurements on the left-hand-side (Cols.~$4-6$) are from \citet{jig02} and errors are 90\% confidence limits. Those on the right-hand-side (Cols.~$7-9$) are from the present work and errors are 68\% confidence limits.}
\label{tbl:herx1_line_list}
\centering
\begin{tabular}{lcc|ccc|ccccccc}
\hline\hline
\noalign{\smallskip} 
Line ID & $\lambda_0$ & State & $\lambda_{\rm obs}$ & $\sigma_v$ & $F_{\rm pho}$ & $\lambda_{\rm obs}$ & $\sigma_v$ & $F_{\rm pho}$ \\
 & $\AA$ & & $\AA$ & ${\rm km~s^{-1}}$ & ${\rm ph~m^{-2}~s^{-1}}$ & $\AA$ & $\AA$ & ${\rm ph~m^{-2}~s^{-1}}$ \\
\noalign{\smallskip} 
\hline
\noalign{\smallskip} 
\ion{Ne}{X} Ly$\alpha$ & 12.134 & a & ... & ... & ... & ... & ... & ... \\
 & & b & ... & ... & $0.7\pm0.3$ & $12.19\pm0.02$ & $1300\pm600$ & $0.8\pm0.2$ \\
 & & c & ... & ... & $0.4\pm0.3$ & ... & ... & ... \\
\ion{Ne}{IX} $r$ & 13.447 & a & ... & ... & ... & ... & ... & ... \\
 & & b &... & ... & $<0.5$ & ... & ... & ... \\
 & & c &... & ... & $0.4\pm0.3$ & ... & ... & ... \\
\ion{Ne}{IX} $i$ & 13.552 & a & ... & ... & ... & ... & ... & ... \\
 & & b & ... & ... & $<0.5$ & $13.50\pm0.03$ & $2000\pm500$ & $1.0\pm0.2$ \\
 & & c & ... & ... & $1.2\pm0.4$ & ... & ... & ... \\
\ion{Ne}{IX} $f$ & 13.698 & a & ... & ... & ... & ... & ... & ... \\
 & & b & ... & ... & $<0.16$ & ... & ... & ... \\
 & & c & ... & ... & $<0.6$ & ... & ... & ... \\
\ion{O}{VIII} Ly$\alpha$ & 18.969 & a & $19.05\pm0.05$ & P Cygni? & $\lesssim9$ & $19.12\pm0.02$ & $1200\pm500$ & $6.0\pm1.0$\\
 & & b & $18.96\pm0.01$ & $390\pm200$ & $1.2\pm0.3$ & $18.971\pm0.008$ & $370\pm170$ & $1.15\pm0.13$ \\
 & & c & $18.96\pm0.02$ & $<450$ & $1.4\pm0.3$ & $18.967\pm0.008$ & $320\pm240$ & $1.4\pm0.2$ \\
\ion{O}{VII} $r$ & 21.602 & a & ... & $3200\pm800$ & $3.8\pm0.8$ & ... & ... & ... \\
 & & b & $21.61\pm0.03$ & $<1600$ & $1.0\pm0.4$ & ... & ... & ... \\
 & & c & $21.62\pm0.02$ & ... & $0.8\pm0.4$ & ... & ... & ... \\ 
\ion{O}{VII} $i$ & 21.803 & a & ... & $3200\pm800$ & $16\pm3$ & $21.77\pm0.03$ & $2200\pm400$ & $11.6\pm1.5$ \\
 & & b & $21.79\pm0.01$ & $<400$ & $3.1\pm0.5$ & $21.785\pm0.005$ & $220\pm220$ & $3.0\pm0.3$ \\
 & & c & $21.82\pm0.02$ & $320\pm160$ & $7.0\pm1.0$ & $21.801\pm0.004$ & $250\pm150$ & $5.9\pm0.4$ \\
\ion{O}{VII} $f$ & 22.101 & a & ... & ... & ... & ... & ... & ... \\
 & & b & ... & ... & $<0.16$ & ... & ... & ... \\
 & & c & ... & ... & $<0.16$ & ... & ... & ... \\
\ion{N}{VII} Ly$\alpha$ & 24.781 & a & $24.85\pm0.17$ & 3200 (fixed) & $8.4\pm0.3$ & $24.80\pm0.08$ & $4200\pm1600$ & $12.9\pm2.4$ \\
 & & b & $24.77\pm0.02$ & $<590$ & $1.5\pm0.3$ & $24.785\pm0.005$ & $100\pm160$ & $1.7\pm0.2$ \\
 & & c & $24.77\pm0.02$ & $<590$ & $2.3\pm0.6$ & $24.777\pm0.008$ & $600\pm140$ & $3.2\pm0.3$ \\
\ion{N}{VI} $r$ & 28.787 & a & ... & ... & ... & ... & ... & ... \\
 & & b & $28.77\pm0.01$ & $<260$ & $0.7\pm0.2$ & $28.783\pm0.010$ & $<150$ & $0.8\pm0.2$ \\
 & & c & $28.75\pm0.02$ & $<260$ & $0.9\pm0.5$ & ... & ... & ... \\ 
\ion{N}{VI} $i$ & 29.083 & a & ... & ... & $\lesssim3$ & ... & ... & ... \\
 & & b & $29.07\pm0.01$ & $270\pm100$ & $2.9\pm0.4$ & $29.083\pm0.005$ & $<110$ & $4.0\pm0.3$ \\
 & & c & $29.10\pm0.02$ & $<260$ & $4.0\pm0.7$ & $29.075\pm0.004$ & $220\pm120$ &  $6.0\pm0.4$ \\
\ion{N}{VI} $f$ & 29.534 & a & ... & ... & ... & ... & ... & ... \\
 & & b & ... & ... & $<0.09$ & ... & ... & ... \\
 & & c & ... & ... & $<0.16$ & ... & ... & ... \\
\ion{C}{VI} Ly$\alpha$ & 33.736 & a & ... & ... & ... & ... & ... & ... \\
 & & b & $33.72\pm0.01$ & $<380$ & $0.6\pm0.2$ & $33.73\pm0.02$ & $<2100$ & $0.8\pm0.2$ \\
 & & c & $33.72\pm0.02$ & $<320$ & $1.0\pm0.4$ & $33.73\pm0.02$ & $340\pm190$ & $1.4\pm0.3$ \\
\noalign{\smallskip} 
\hline
\end{tabular}
\end{table*}

\end{document}